\documentclass{article}
\usepackage{amssymb}
\usepackage{amsmath}

\usepackage{graphics}

\input{tcilatex}
\begin{document}

\title{Maximum Entropy approach to a Mean Field Theory for Fluids{\small 
\thanks{%
Presented at MaxEnt 2002, the 22th International Workshop on Bayesian
Inference and Maximum Entropy Methods (August 3-7, 2002, Moscow, Idaho, USA)}%
}}
\author{Chi-Yuan Tseng{\small \thanks{%
E-mail: ct7663@csc.albany.edu}} and Ariel Caticha{\small \thanks{%
E-mail: ariel@albany.edu}} \\
{\small Department of Physics, University at Albany-SUNY, }\\
{\small Albany, NY 12222, USA.}}
\date{}
\maketitle

\begin{abstract}
Making statistical predictions requires tackling two problems: one must
assign appropriate probability distributions and then one must calculate a
variety of expected values. The method of maximum entropy is commonly used
to address the first problem. Here we explore its use to tackle the second
problem. We show how this use of maximum entropy leads to the Bogoliuvob
variational principle which we generalize, apply to density functional
theory, and use it to develop a mean field theory for classical fluids.
Numerical calculations for Argon gas are compared with experimental data.
\end{abstract}

\section{Introduction}

The method of Maximum Entropy (ME) has been designed to solve the general
problem of updating from a prior probability distribution to a posterior
distribution when new information in the form of a constraint becomes
available \cite{footnote1}. The method reflects the deep conviction that one
should not change one's mind frivolously, that whatever was learned in the
past is important. Indeed, this concern for prior information is the only
guarantee that the new information itself will not be easily discarded in
the future. The chosen posterior distribution should coincide with the prior
as closely as possible: one should only update those aspects of one's
beliefs for which hard new evidence has been supplied \cite{Jaynes57}-\cite%
{Caticha00}.

The purpose of this work is to explore the use of the ME method as a means
to tackle problems of a very different kind. The assignment of a probability
distribution that faithfully codifies our state of knowledge is only a first
step towards answering our questions about a physical system. There is a
second and crucial step that requires us to extract the desired answers from
the probability distribution. It is not enough to write down a probability
distribution, we must also be able to read it, to figure out what it means,
what it implies. We must be able to calculate the expected values of
whatever quantities happen to be of interest to us. This is a difficult
practical problem for which no solution of universal applicability is known
and, accordingly, a wide variety of different techniques, both analytical
and numerical, have been developed.

The approach we explore here consists of approximating the `exact'
distribution by a more tractable one. Then, the question is: given a family
of trial tractable distributions, which one should we select? Which
distribution within the given family best approximates the exact one? When
phrased in this way it is clear that this is the kind of question that can
be tackled using the method of ME: one must select the trial distribution
that reflects the least misrepresentation of our state of knowledge. More
specifically, the chosen distribution should be that which maximizes the
appropriate relative entropy.

The purpose of this work is to explore the use of ME as a method to generate
approximations. As a simple illustration, in section 2, we apply the ME
method to canonical Gibbs-Boltzmann distributions. The resulting variational
technique, when expressed in terms of the free energies, is recognized as
the Bogoliuvob variational principle \cite{Callen85}. This is a well known
technique for generating mean field approximations for discrete systems on a
lattice. What is perhaps not as widely known is that the Bogoliuvob
principle is just another application of the ME method (however, see \cite%
{Opper01}).

In the bulk of the paper we explore the use of the ME method to generate
mean field (MF) approximations for classical fluids. Ever since the
pioneering work of van der Waals many different versions of MF theories have
been proposed and it is not always clear what they have in common -- except
perhaps for neglecting correlation effects. It appears that none of these
versions have been derived as a systematic application of the ME method.

The study of classical fluids is an old and mature field. There exist
extensive treatments in many excellent books and reviews \cite{Hansen86}.
Our goal here is not so much to contribute to the study of the fluids
themselves but rather to use this well-explored but still non-trivial field
as a testing ground for further development of the ME method. In future work
we intend to build on our current results and extend the method to generate
other more sophisticated approximations involving canonical transformations
to collective coordinates; only then do we expect to obtain significantly
new contributions to the theory of fluids.

The study of both homogeneous and inhomogeneous fluids is currently best
approached using density functional methods \cite{Hohenberg64}\cite{Ebner76}%
\cite{Evans92} which have been noted to show a remarkable formal similarity
to statistical thermodynamics \cite{Argaman00}. In section 3 we show that
the resemblance goes beyond a mere accident: density functional methods
constitute a generalization of the Gibbs grand canonical ensemble to the
case of a spatially dependent chemical potential. The resulting `density'
distribution is, however, too complicated for actual calculations and
approximations are needed. In section 4 the MF approximation for the
density, various thermodynamic potentials and pair distribution functions is
derived. Following an idea due to Percus \cite{Percus62}, a second, somewhat
improved MF estimate of the two-particle distribution of a homogeneous fluid
is obtained by expressing it in terms of the one-particle distribution of
the fluid placed in a suitable external potential. An explicit comparison
with experimental data for Argon is carried out in section 5. This serves
both to determine the parameters of the Lennard-Jones potential used to
model the interaction between Argon atoms and to explore the limitations of
the MF approximation. Conclusions and final comments appear in section 6.

\section{Maximum Entropy as an approximation technique}

Let the microstates of a system (for example, its location in phase space or
perhaps the values of spin variables) be labelled by coordinates $q$, and
let us assume that the probability that the system is in a microstate within
a particular range $dq$ is given by the canonical distribution 
\begin{equation}
P(q)dq=\frac{e^{-\beta H(q)}}Zdq\,\quad \text{where}\quad Z=e^{-\beta
F}=\int dq\,e^{-\beta H(q)}.  \label{P(q)}
\end{equation}
In principle knowing the distribution $P$ we should be able to answer any
question we might care to ask but in practice complications arise. Any
realistic Hamiltonian $H(q)$ will usually include complicated non linear
interactions that make the task of calculating expected values or most other
integrals over $P$ impossibly difficult.

A possible way around this problem consists of replacing the `exact'
distribution $P$ by another more tractable distribution $P_0$ that
approximately represents the same state of knowledge. This is perhaps too
much to ask for. What we really want is a distribution $P_0$ that
approximately represents those aspects of the information in $P$ that happen
to be relevant to those very few, very specific questions that we are
actually asking.

The practical problem of choosing a $P_0$ is dealt with in two steps. First
we search for a family of trial distributions that are simple enough that
actual calculations are feasible and that we have some reason to suspect
might codify the appropriate relevant information. This step is the
difficult one because there is no known systematic procedure to carry it
out; it is a matter trial and error guided by intuition. The second and
easier step is to select the one member within the trial family that best
resembles the exact $P$. This step is easier because there exists a
systematic, mechanical procedure to be followed. As discussed in \cite%
{Caticha00} (and other references therein) there is a unique selection
criterion satisfying natural desiderata of consistency and objectivity: the
distribution to be selected is that which maximizes the entropy of $P_0$
relative to $P$, 
\begin{equation}
S[P_0,P]=-\int dq\,P_0(q)\log \frac{P_0(q)}{P(q)}\,.  \label{relat S}
\end{equation}

The success of the whole method hinges crucially on the first step, the
choice of the family of trial distributions. To increase the likelihood that
important relevant features (\emph{e.g.}, symmetries or dominant
interactions) are not left out it is usual -- but not necessary -- to select
trial distributions that could conceivably model an idealized system. Such
trial distributions are also canonical but with a modified Hamiltonian $%
H_0(q;\alpha )$ where $\alpha $ are parameters labeling each distribution
within the family, 
\begin{equation}
P_0(q;\alpha )dq=\frac{e^{-\beta H_0(q;\alpha )}}{Z_0}dq\,\,\quad \text{where%
}\quad Z_0(\alpha )=e^{-\beta F_0(\alpha )}=\int dq\,e^{-\beta H_0(q;\alpha
)}.  \label{P0(q)}
\end{equation}
Substituting into eq.(\ref{relat S}) gives, 
\begin{equation}
S[P_0,P]=\beta \left( \langle H_0-H\rangle _0-F_0+F\right) \,,
\end{equation}
where $\langle \ldots \rangle _0$ refers to averages over the trial $P_0$.
The inequality $S[P_0,P]\leq 0$, can then be written as 
\begin{equation}
F\leq F_0+\langle H-H_0\rangle _0\,.
\end{equation}
Thus, one selects $P_0(q;\alpha )$ by maximizing $S$ over all values of $%
\alpha $ or equivalently by minimizing the quantity $F_0+\langle
H-H_0\rangle _0$. This alternate form of the variational principle and its
use to generate approximations is well known. It is usually associated with
the name of Bogoliuvob \cite{Callen85} and it is the main technique to
generate MF approximations for discrete systems of spins on a lattice. What
is perhaps not as widely known is that the Bogoliuvob variational principle
is just a peculiar application of the ME method.

After this brief formal illustration we are ready to tackle the problem of
fluids.

\section{The canonical `density' ensemble}

Many of the most interesting questions we ask when discussing fluids can be
addressed once we know the density of the fluid under various experimental
conditions. Examples include the transitions between the gas and liquid
phase, the structure of the gas-liquid interface, surface tension and
capillarity effects and so on. We therefore would like to have a formalism
where the (possibly nonuniform) density $n(r)$ appears as an explicit
variable. It is very easy to do this, just find the probability distribution
that maximizes entropy subject to constraints on the expected energy and on
the expected density. The resulting distribution is a generalized canonical
ensemble we will call the `density' ensemble.

The logic of the previous paragraph may sound unfamiliar and require further
comments. When justifying the use of the ME method to obtain the canonical
Boltzmann-Gibbs distribution ($P_q\propto e^{-\beta E_q}$) one might say
something like ``we seek the minimally biased (\emph{i.e.} maximum entropy)
distribution that codifies the information we do possess (the expected
energy) and nothing else''. Many authors -- even those who fully appreciate
the value of the concept of entropy in inductive reasoning -- find this
justification unsatisfactory. Indeed, they would argue, for example, that
the spectrum of black body radiation is what it is independently of our
knowledge. We prefer to phrase the objection differently: in most realistic
situations the one thing that is not known is the expected energy, and even
then one still makes correct predictions by maximizing entropy subject to a
constraint on the (unknown) expected energy. But then, how can we justify
the imposition of such `unknown' constraints in the ME\ method?

We propose here that the motivation behind imposing a constraint on the
expected energy should not be that this is a quantity we happen to know --
because we do not -- but rather that we recognize the\emph{\ }expected
energy as the quantity we \emph{need }to know. We recognize that in many
experimental situations the expected energy, even if unknown, is the crucial
quantity that codifies the relevant information and we proceed to theorize
as if we knew it. All resulting predictions will contain the temperature as
a free unknown parameter. Later, when it comes to actually comparing
theoretical predictions with experimental data we expect that part of the
data analysis will consist of adjusting the unknown temperature parameter to
provide the best fit to the data. In other words, the information on
expected energy or temperature is something to be obtained from the
experiment itself.

Thus, the motivation behind imposing a constraint on the density is not that
it is something we know, but rather that it is what we should know; it is
the quantity that codifies information that is very relevant for most
questions of interest and therefore it is important that it appear
explicitly in the formalism.

The probability that the fluid is composed of $N$ particles with positions
and momenta within the phase space volume $dq_N$ at the location $%
q_N=\{p_i,r_i;\;i=1,...,N\}$ is given by 
\begin{equation}
P(q_N;\beta ,\lambda )\,dq_N
\end{equation}
where 
\begin{equation}
dq_N=\frac 1{N!\,h^{3N}}\prod_{i=1}^Nd^3p_id^3r_i\,,
\end{equation}
and 
\begin{equation}
P(q_N;\beta ,\lambda )=\frac 1Z\exp -\beta \left[ H(q_N)+\int
d^3r\,\,\lambda (r)\hat{n}(r)\right] \,,  \label{exact dist}
\end{equation}
where $\lambda (r)$ are Lagrange multipliers that enforce the constraint on
the expected density $\langle \hat{n}(r)\rangle $ at each point in space and
the density is 
\begin{equation}
\hat{n}(r)=\sum_{i=1}^N\,\delta (r-r_i)\,.
\end{equation}
$H(q_N)$ is the Hamiltonian, 
\begin{equation}
H(q_N)=\sum_{i=1}^N\,\frac{p_i^2}{2m}+\sum_{i>j}^Nu(r_{ij})+%
\sum_{i=1}^Nv(r_i)\,,
\end{equation}
where $r_{ij}=r_i-r_j$, $u(r_{ij})$ is the two-body interparticle potential,
and $v(r)$ is an external potential. The partition function is 
\begin{equation}
Z(T,\lambda )=\sum_{N=0}^\infty \int dq_N\,\exp -\beta \left[ H(q_N)+\int
d^3r\,\,\lambda (r)\hat{n}(r)\right] \,\overset{\limfunc{def}}{=}e^{-\beta
\Omega (T,\lambda )}\,,
\end{equation}
where we introduced the thermodynamic potential $\Omega (T,\lambda )$, a
function of the temperature $T\equiv 1/\beta $ (units are such that $k_B=1$%
), and a functional of $\lambda (r)$.

The interparticle and the external potential energies can be rewritten in
terms of the density, 
\begin{equation}
\sum_{i=1}^Nv(r_i)=\sum_{i=1}^N\int d^3r\,\,v(r)\,\delta (r-r_i)=\int
d^3r\,\,v(r)\hat{n}(r)\,,
\end{equation}
and 
\begin{equation}
U=\sum_{i>j}^Nu(r_{ij})=\frac 12\sum_{i\neq j}^Nu(r_{ij})=\frac 12\int
d^3rd^3r^{\prime }\,\,u(r-r^{\prime })\hat{n}^{(2)}(r,r^{\prime }),\,
\label{U}
\end{equation}
where $\hat{n}^{(2)}(r,r^{\prime })$ is the two-particle distribution, 
\begin{equation}
\hat{n}^{(2)}(r,r^{\prime })=\sum_{i\neq j}\,\delta (r-r_i)\,\delta
(r^{\prime }-r_j)=\hat{n}(r)\hat{n}(r^{\prime })-\hat{n}(r)\delta
(r-r^{\prime })\;.  \label{n2}
\end{equation}

We remark that a constraint on $\langle \hat{n}(r)\rangle $ also constrains
the expected number of particles 
\begin{equation}
\langle N\rangle =\int d^3r\,\langle \hat{n}(r)\rangle \,.
\end{equation}
Therefore the density ensemble is a generalization of the grand-canonical
ensemble. Indeed, when $\lambda (r)$ is a constant the density distribution
reduces to the grand-canonical distribution and $-\lambda $ coincides with
the chemical potential $\mu $.

Notice also that the effects of the Lagrange multiplier field $\lambda (r)$
are indistinguishable from the effects of the external potential $v(r)$.
Similarly, when two-particle correlations $\langle \hat{n}^{(2)}(r,r^{\prime
})\rangle $ carry relevant information we might want to develop a formalism
where they appear explicitly. The effects of the corresponding Lagrange
multipliers are indistinguishable from those of the two-body potentials $%
u(r,r^{\prime })$. Perhaps this is the ultimate reason why potentials in
general are valuable in physics: they establish control over the relevant
physical quantities, the density, and the two-particle correlations.

The expected value of the density is 
\begin{equation}
\langle \hat{n}(r)\rangle =\frac{\delta \Omega }{\delta \lambda (r)}\overset{%
\limfunc{def}}{=}n(r)\,,
\end{equation}
and the entropy is 
\begin{equation}
S[P]=-\sum_N\int dq_N\,P\log P=-\frac{\partial \Omega }{\partial T}
\end{equation}
These two equations can be combined into 
\begin{equation}
d\Omega =-SdT+\int d^3r\,n(r)\delta \lambda (r)\,.
\end{equation}

To achieve a formulation in terms of the density $n(r)$ rather than the
multiplier $\lambda (r)$ we consider the Legendre transform 
\begin{equation}
\Phi [T,n]=\Omega [T,\lambda ]-\int d^3r\,\,\lambda (r)n(r)\,,
\end{equation}
so that

\begin{equation}
d\Phi =-SdT-\int d^3r\,\lambda (r)\delta n(r)\,.
\end{equation}

The density functional formalism of Hohenberg and Kohn \cite{Hohenberg64} is
founded upon a theorem proving the existence of a functional $\Phi [T,n]$
that is independent of the external potential $v(r)$. From the perspective
afforded by the ME approach the independence of $\Phi [T,n]$ on the
potential $v(r)$ is a triviality achieved by construction -- through the
Legendre transform -- and their truly nontrivial insight is the recognition
of the density as the correct choice of relevant variable.

\section{Mean field approximation}

\subsection{MF trial distributions}

We want to approximate the `exact' probability distribution by a more
tractable trial distribution in which the interparticle interactions $U$ are
replaced by an interaction with an external potential, the mean field. The
problem consists of selecting the best trial distribution, that which best
approximates the exact distribution. The drastic approximation being made is
the neglect of two-particle correlations induced by the interparticle
potential.

The trial MF distribution is 
\begin{equation}
P_0(q_N;\beta ,\lambda )=\frac 1Z\exp -\beta \left[ H_0(q_N)+\int
d^3r\,\,\lambda (r)\hat{n}(r)\right] \,,  \label{trial P a}
\end{equation}
where 
\begin{equation}
H_0(q_N)=\sum_{i=1}^N\,\frac{p_i^2}{2m}+\int d^3r\,\,\left(
v(r)+v_0(r)\right) \hat{n}(r)\,,
\end{equation}
where $v_0(r)$ is the mean field to be determined.

First we compute several thermodynamic quantities of interest. It is
convenient to absorb the external potential $v(r)$, the mean field $v_0(r)$,
and the multiplier field $\lambda (r)$ into a single potential $V(r)$, 
\begin{equation}
V(r)=v(r)+v_0(r)+\lambda (r)\,.  \label{V}
\end{equation}
The partition function $Z_0$ is 
\begin{equation}
Z_0=\sum_{N=0}^\infty \frac 1{N!}\left[ \int \frac{d^3p}{h^3}\,\exp -\frac{%
\beta p^2}{2m}\right] ^N\left[ \int d^3r\,\exp -\beta V(r)\right] ^N\,%
\overset{\limfunc{def}}{=}e^{-\beta \Omega _0(\beta ,\lambda )}\,,
\end{equation}
so that 
\begin{equation}
\Omega _0[T,\lambda ]=-\frac 1{\beta \Lambda ^3}\int d^3r\,\,e^{-\beta
V(r)}\quad \text{where}\quad \Lambda =\left( \frac{\beta h^2}{2\pi m}\right)
^{1/2}\,.  \label{Omega0}
\end{equation}
The expected density is 
\begin{equation}
\langle \hat{n}(r)\rangle _0=\frac{\delta \Omega _0}{\delta \lambda (r)}=%
\frac{\delta \Omega _0}{\delta V(r)}\overset{\limfunc{def}}{=}n_0(r)
\label{dOmega/dV}
\end{equation}
or 
\begin{equation}
n_0(r)=\frac{e^{-\beta V(r)}}{\Lambda ^3}\,.  \label{n0}
\end{equation}

\subsection{ME optimization of the mean field}

Within the family of trial MF distributions the one that is closest to the
canonical density ensemble is that which maximizes the relative entropy 
\begin{equation}
S[P_0|P]=-\sum_{N=0}^\infty \int dq_N\,P_0(q_N|\beta ,\lambda ,v_0)\log 
\frac{P_0(q_N|\beta ,\lambda ,v_0)}{P(q_N|\beta ,\lambda )}
\end{equation}
Substituting Eqs.(\ref{exact dist}) and (\ref{trial P a}) we obtain 
\begin{equation}
S[P_0|P]=\beta \left[ \Omega -\Omega _0-\langle H-H_0\rangle _0\right] \,.
\end{equation}
(The subscript $0$ in $\langle \cdots \rangle _0$ indicates the averages are
computed over the trial distribution $P_0$.) Since $S[P_0|P]\leq 0$, we have 
\begin{eqnarray}
\Omega [T,\lambda ] &\leq &\Omega _U[T,\lambda ,v_0]\overset{\limfunc{def}}{=%
}\Omega _0+\langle H-H_0\rangle _0 \\
&=&\Omega _0+\langle U\rangle _0-\int d^3r\,v_0(r)n_0(r)\,,  \label{OmegaU}
\end{eqnarray}
and maximizing $S[P_0|P]$ is equivalent to minimizing $\Omega _U$ over all
mean fields $v_0.$ The MF approximation to $\Omega $, denoted $\bar{\Omega}$%
, is defined as the best $\Omega _U$, 
\begin{equation}
\Omega [T,\lambda ]\approx \bar{\Omega}[T,\lambda ]\overset{\limfunc{def}}{=}%
\min_{v_0}\Omega _U[T,\lambda ,v_0]\,.
\end{equation}
To calculate the potential energy $\langle U\rangle _0$ use Eqs.(\ref{U})
and (\ref{n2}) 
\begin{equation}
\langle U\rangle _0=\frac 12\int d^3rd^3r^{\prime }\,\,u(r-r^{\prime
})n_0^{(2)}(r,r^{\prime })\,,
\end{equation}
where 
\begin{equation}
n_0^{(2)}(r,r^{\prime })=\langle \hat{n}^{(2)}(r,r^{\prime })\rangle
_0=\langle \hat{n}(r)\hat{n}(r^{\prime })\rangle _0-n_0(r)\delta
(r-r^{\prime })\,,
\end{equation}
furthermore, from Eq.(\ref{trial P a}) and (\ref{Omega0}-\ref{n0}) 
\begin{equation}
\langle \hat{n}(r)\hat{n}(r^{\prime })\rangle _0=n_0(r)n_0(r^{\prime
})+n_0(r)\delta (r-r^{\prime })\,.
\end{equation}
Therefore 
\begin{equation}
n_0^{(2)}(r,r^{\prime })=n_0(r)n_0(r^{\prime })\,,
\end{equation}
and the result for $\langle U\rangle _0$ is 
\begin{equation}
\langle U\rangle _0=\frac 12\int d^3rd^3r^{\prime }\,\,u(r-r^{\prime
})n_0(r)n_0(r^{\prime })\,.
\end{equation}

The best choice of mean field $v_0(r)$ is given by 
\begin{equation}
0=\frac{\delta \Omega _U}{\delta v_0(r)}=\frac{\delta \Omega _U}{\delta V(r)}%
=\beta n_0(r)\left[ v_0(r)-\int d^3r^{\prime }\,\,u(r-r^{\prime
})n_0(r^{\prime })\right]
\end{equation}
where the multiplier $\lambda (r)$ and the external field $v(r)$ are assumed
fixed and we used Eqs.(\ref{OmegaU}) and (\ref{n0}). Therefore, the best
mean field, which we will denote by $\bar{v}(r)$ is a function of the
temperature $T$ and a functional of the multiplier $\lambda (r)$ and the
external field $v(r)$, 
\begin{equation}
\bar{v}=\bar{v}[T,\lambda ,v]\,,
\end{equation}
defined by the equation 
\begin{equation}
\bar{v}(r)-\int d^3r^{\prime }\,\,u(r-r^{\prime })\bar{n}_0(r)=0\,,
\label{vbar}
\end{equation}
where $\bar{n}_0$ is a function of $\bar{v}$, 
\begin{equation}
\bar{n}_0(r)\overset{\limfunc{def}}{=}\frac{e^{-\beta \bar{V}(r)}}{\Lambda ^3%
}\quad \text{and}\quad \bar{V}(r)\overset{\limfunc{def}}{=}v(r)+\bar{v}%
(r)+\lambda (r)\,\,.  \label{vbar2}
\end{equation}

To summarize, the MF approximation to the exact thermodynamic potential is
given by 
\begin{equation}
\bar{\Omega}[T,\lambda ,v]=\bar{\Omega}_0+\bar{U}_0-\int d^3r\,\bar{v}(r)%
\bar{n}_0(r)  \label{Omegabar}
\end{equation}
where 
\begin{equation}
\bar{\Omega}_0=-\frac 1\beta \int d^3r\,\,\bar{n}_0(r)\,,
\end{equation}
\begin{equation}
\bar{U}_0=\frac 12\int d^3rd^3r^{\prime }\,\,u(r-r^{\prime })\bar{n}_0(r)%
\bar{n}_0(r^{\prime })\,,
\end{equation}
and $\bar{v}(r)$ is given by Eq.(\ref{vbar}). Throughout we will use
overbars to denote quantities evaluated in the MF approximation.

Notice that in replacing the exact distribution $P$ by the approximate $P_0$
we are not quite replacing the fluid described by the Hamiltonian $H$ by a
different fluid described by $H_0$: we compute the expectation of the exact $%
H$ in the approximate $P_0$ rather than the expectation of the approximate $%
H_0$ in the approximate $P_0$.

\subsection{Density in the MF approximation}

Entropy and density are obtained as derivatives of the thermodynamic
potential,

\begin{equation}
d\bar{\Omega}=-\bar{S}dT+\int d^3r\,\bar{n}(r)\delta \lambda (r)\,.
\label{dOmegabar}
\end{equation}
Using Eq.(\ref{vbar}) and (\ref{vbar2}) the density $n(r)$ is given by 
\begin{equation}
\bar{n}(r)=\frac{\delta \bar{\Omega}}{\delta \lambda (r)}=\bar{n}_0(r)\,=%
\frac{e^{-\beta \bar{V}(r)}}{\Lambda ^3}\,.  \label{nbar}
\end{equation}
This permits us to interpret Eq.(\ref{vbar}) as a self consistency equation:
the mean field $\bar{v}(r)$ is generated by a molecular distribution
described by the density $\bar{n}(r)$, while the molecules distribute
themselves according to the mean field.

The Legendre transform of $\bar{\Omega}[t,\lambda ]$, 
\begin{equation}
\bar{\Phi}[T,\bar{n}]=\bar{\Omega}[T,\lambda ]-\int d^3r\,\,\lambda (r)\bar{n%
}(r)\,.
\end{equation}
gives the MF approximation to the density functional, 
\begin{eqnarray}
\bar{\Phi}[T,\bar{n}] &=&\int d^3r\,\,\bar{n}(r)\left[ T\left( \log \Lambda
^3\bar{n}(r)-1\right) +v(r)\right]  \nonumber \\
&&+\frac 12\int d^3rd^3r^{\prime }\,\,u(r-r^{\prime })\bar{n}(r)\bar{n}%
(r^{\prime })\,.  \label{Phibar}
\end{eqnarray}
The density $\bar{n}(r)$ in a given external potential and multiplier fields
is determined by 
\begin{equation}
\frac{\delta \bar{\Phi}}{\delta \bar{n}(r)}=-\lambda (r)\,,
\end{equation}
or 
\begin{equation}
\log \Lambda ^3\bar{n}(r)=-\beta \left[ v(r)+\lambda (r)+\int d^3r^{\prime
}\,\,u(r-r^{\prime })\bar{n}(r^{\prime })\right] \,,
\end{equation}
which is clearly equivalent to Eqs.(\ref{vbar}) and (\ref{vbar2}).

If the only constraint on the density is a constraint on the expected total
number of particles then the multiplier field $\lambda (r)$ is a constant
which we set equal to minus the chemical potential. Therefore, 
\begin{equation}
T\log \Lambda ^3\bar{n}(r)+v(r)+\int d^3r^{\prime }\,\,u(r-r^{\prime })\bar{n%
}(r^{\prime })=\mu \,.  \label{constant lambda}
\end{equation}
We can either consider this equation as determining the chemical potential
given the density, or alternatively, it determines the density for a given
chemical potential.

\subsection{Correlation functions -- General}

First we recall some standard definitions. Density correlation functions are
given by derivatives of the partition function $Z[T,\lambda ,v]$ with
respect to $\lambda (r_i)$. Derivatives of the thermodynamic potential $%
\Omega [T,\lambda ,v]$ yield the so called `connected' correlation
functions; the first is the expected density itself, 
\begin{equation}
n(r)\overset{\func{def}}{=}\langle \hat{n}(r)\rangle =\frac{\delta \Omega }{%
\delta \lambda (r)}\,.
\end{equation}
The second is the density fluctuation correlation, 
\begin{eqnarray}
&&G^{(2)}(r_1,r_2)\overset{\func{def}}{=}\left\langle \left( \hat{n}%
(r_1)-n(r_1)\right) \left( \hat{n}(r_2)-n(r_2)\right) \right\rangle 
\nonumber \\
&=&\frac{-1}\beta \frac{\delta ^2\Omega }{\delta \lambda (r_1)\delta \lambda
(r_2)}=\frac{-1}\beta \frac{\delta n(r_1)}{\delta \lambda (r_2)}
\label{dens fluct}
\end{eqnarray}
Using 
\begin{equation}
n^{(2)}(r_1,r_2)=\langle \hat{n}^{(2)}(r_1,r_2)\rangle =\langle \hat{n}(r_1)%
\hat{n}(r_2)\rangle -n(r_1)\delta (r_1-r_2)\,,
\end{equation}
$G^{(2)}$ can be reexpressed as 
\begin{equation}
G^{(2)}(r_1,r_2)=n^{(2)}(r_1,r_2)+n(r_1)\delta (r_1-r_2)-n(r_1)n(r_2)\,.
\label{dens fluct 2}
\end{equation}

To introduce the direct (or one-particle irreducible) correlations define $%
\Gamma ^{(2)}$ as the inverse of $G^{(2)}$, $\Gamma ^{(2)}=[G^{(2)}]^{-1}$, 
\begin{equation}
\int d^3r_3G^{(2)}(r_1,r_3)\Gamma ^{(2)}(r_3,r_2)=\delta (r_1-r_2)\,.
\label{OZ1}
\end{equation}
Therefore, using Eq.(\ref{dens fluct}) and the chain rule, 
\begin{equation}
\Gamma ^{(2)}(r_1,r_2)=-\beta \frac{\delta \lambda (r_1)}{\delta n(r_2)}%
=\beta \frac{\delta ^2\Phi }{\delta n(r_1)\delta n(r_2)}\,.  \label{Gamma2}
\end{equation}

From Eq.(\ref{Phibar}) we see that for an ideal gas we have 
\begin{equation}
\Gamma _{\func{id}}^{(2)}(r_1,r_2)=\frac{\delta (r_1-r_2)}{n(r_1)}\,.
\label{Gamma2id}
\end{equation}
The difference between $\Gamma ^{(2)}$ and $\Gamma _{\func{id}}^{(2)}$ is
attributed to the interactions and defines the direct correlation function $%
c $, 
\begin{equation}
\Gamma ^{(2)}(r_1,r_2)=\frac{\delta (r_1-r_2)}{n(r_1)}-c(r_1,r_2)\,.
\label{Gamma2c}
\end{equation}

It is convenient to introduce the pair-distribution function $g(r_1,r_2)$ 
\begin{equation}
n^{(2)}(r_1,r_2)=n(r_1)n(r_2)g(r_1,r_2)\,.  \label{g(r)}
\end{equation}
Substituting this and Eq.(\ref{dens fluct 2}) into Eq.(\ref{OZ1}) gives 
\begin{equation}
g(r_1,r_2)=1+c(r_1,r_2)+\int d^3r_3n(r_3)c(r_3,r_2)\left[ g(r_1,r_3)-1\right]
\,.  \label{OZ}
\end{equation}
This is called the Oernstein-Zernicke equation. Notice that there is no
physical content to the OZ equation; it follows exactly from the definitions
of $\Gamma ^{(2)}$, $c$ and $g$. This concludes our brief summary of
definitions.

At high fluid densities there is data from x-ray and neutron diffraction
experiments that gives the radial distribution $g(r)$ directly. At low
densities the experimental data is less direct. The data that will be
discussed below is derived from the virial expansion to the equation of
state, in particular the second virial coefficient, and from measurements of
the internal energy of the fluid.

For a homogeneous, isotropic fluid of bulk density $N/V=n_b$ the equation of
state is given in terms of $g(r)$ by \cite{Hansen86} 
\begin{equation}
\frac P{n_bkT}=1-\frac{n_b}{6kT}\int d^3r\,r\frac{du\left( r\right) }{dr}%
g\left( r\right) \,.  \label{eq of state}
\end{equation}
Using eq.(\ref{U}) and (\ref{g(r)}) the internal energy is 
\begin{equation}
E=\frac 32NkT+\langle U\rangle
\end{equation}
where, setting the external potential $v(r)$ equal to zero, the potential
energy per particle is 
\begin{equation}
\frac{\langle U\rangle }N=\frac{n_b}2\int d^3r\,u(r)g\left( r\right) \,.
\label{U per part}
\end{equation}

\subsection{MF approximation to $g(r)$}

The MF approximation to the density-fluctuation correlation is obtained
using Eqs.(\ref{dens fluct}) and (\ref{nbar}), 
\begin{equation}
\bar{G}^{(2)}(r_1,r_2)=\,\bar{n}(r_1)\left[ \frac{\delta \bar{v}(r_1)}{%
\delta \lambda (r_2)}+\delta (r_1-r_2)\right] \,\,.
\end{equation}
The functional derivative on the right is obtained from the selfconsistency
equation defining the mean field $\bar{v}$, Eq.(\ref{vbar}). The expression
for $\bar{\Gamma}^{(2)}$ is much simpler and can be written in closed form, 
\begin{equation}
\bar{\Gamma}^{(2)}(r_1,r_2)=\frac{\delta (r_1-r_2)}{n(r_1)}+\beta
u(r_1-r_2)\,.
\end{equation}
Therefore the direct correlation function is 
\begin{equation}
\bar{c}(r_1,r_2)=-\beta u(r_1-r_2)\,.  \label{cbar}
\end{equation}
Using this expression for $\bar{c}$ in the case of a homogeneous, isotropic
fluid of bulk density $n_b$ in the OZ equation (\ref{OZ}), gives 
\begin{equation}
\bar{g}(r)=1-\beta u(r)-\beta n_b\int d^3r^{\prime }u(r^{\prime })\left[ 
\bar{g}(r-r^{\prime })-1\right] \,,  \label{gbar }
\end{equation}
where we have set $r_2=0$. This is the integral equation that must be solved
to obtain $\bar{g}(r)$ for each $T$ and $n_b$.

\subsection{An improved MF approximation to $g(r)$}

The approximation in which each particle moves in a MF field generated by
all the others is too drastic. One can obtain a somewhat better
approximation -- still of the MF type -- by using an elegant argument due to
Percus \cite{Percus62}. The strategy is to write multiparticle distribution
functions for a fluid in terms of the single particle density of the same
fluid placed in a suitable external potential.

The idea is simple and revolves around the product rule for probabilities.
The probability $P(r)d^3r$ that a particle is found within $d^3r$ at $r$ is
obtained from Eq.(\ref{exact dist}) by marginalizing over the remaining $N-1$
particles, 
\begin{equation}
P(r)=\sum_i\,\langle \delta (r-r_i)\rangle \overset{\func{def}}{=}n(r|v)\,,
\end{equation}
where the average is taken with the exact distribution, Eq.(\ref{exact dist}%
), and for later convenience the dependence on the external potential $v$ is
indicated explicitly. Similarly, the probability that one particle is found
at $r$ and a second one at $r_s$ is given by $P^{(2)}(r,r_s)d^3rd^3r_s$ with

\begin{equation}
P(r,r_s)=\langle \sum_i\,\delta (r-r_i)\sum_{j\neq i}\,\delta
(r_s-r_j)\rangle \overset{\func{def}}{=}n^{(2)}(r,r_s|v).
\label{two particle}
\end{equation}
The product rule gives 
\begin{equation}
P(r,r_s)=P(r_s)P(r|r_s)\,,
\end{equation}
but 
\begin{equation}
P(r|r_s)=N\langle \sum_{i=1}^{N-1}\,\delta (r-r_i)\,\delta (r_s-r_N)\rangle
\,\overset{\func{def}}{=}n(r|v_s)\,
\end{equation}
is exactly the one-particle distribution of a fluid in a modified external
potential, $v_s(r)=v(r)+u(r-r_s)$. Therefore, 
\begin{equation}
n^{(2)}(r,r_s|v)=n(r_s|v)n(r|v_s)\,,
\end{equation}
and using 
\begin{equation}
n^{(2)}(r,r_s|v)=n(r|v)n(r_s|v)g(r,r_s|v)\,,
\end{equation}
we get 
\begin{equation}
g(r,r_s|v)=\frac{n(r|v_s)}{n(r|v)}\,.  \label{Percus}
\end{equation}
This is exact and can be trivially generalized to multiparticle
distributions.

Now we return to the MF\ approximation. If there is no external potential, $%
v(r)=0$ the fluid is homogeneous and isotropic with a density $n_b$. We
single out one molecule located at $r_s$ and treat it as special. Since the
system is homogeneous $n(r|v=0)=n_b$. The remaining molecules are
effectively moving in the external potential $v_s(r)=u(r-r_s)$ due to the
special molecule. The density $\bar{n}(r|v_s)$ at location $r$ is given by
Eq.(\ref{constant lambda}),

\begin{equation}
T\log \Lambda ^3\bar{n}(r|v_s)=\mu -u(r-r_s)-\int d^3r^{\prime
}\,\,u(r-r^{\prime })\bar{n}(r^{\prime }|v_s)\,.
\end{equation}
On the other hand, given $n_b$, the chemical potential is determined from a
similar equation with the external potential set to zero,

\begin{equation}
T\log \Lambda ^3n_b=\mu -\int d^3r^{\prime }\,\,u(r-r^{\prime })n_b.
\end{equation}
Subtracting these two equations gives

\begin{equation}
T\log \frac{\bar{n}(r|v_s)}{n_b}=-u(r-r_s)-n_b\int d^3r^{\prime
}\,\,u(r-r^{\prime })\left[ \frac{\bar{n}(r^{\prime }|v_s)}{n_b}-1\right] \,.
\end{equation}
Therefore, setting $r_s=0$, the MF approximation to Eq.(\ref{Percus}) is

\begin{equation}
\bar{g}_s(r)=\exp \left( -\beta u(r)-\beta n_b\int d^3r^{\prime
}\,\,u(r^{\prime })\left[ \bar{g}_s(r-r^{\prime })-1\right] \right) \,.
\label{gbars}
\end{equation}

We have described two MF approximations to $g(r|v)$. The first requires
solving Eq.(\ref{gbar }) for $\bar{g}(r)$. The second is an improved MF
approximation, which does include some correlation effects, is obtained by
solving Eq.(\ref{gbars}) for $\bar{g}_s(r)$.

\section{An explicit example: Argon}

Realistic model interatomic potentials must include a long range weak
attraction and a short range strong repulsion if they are to predict the
properties of dense fluids. A popular but by no means unique choice for
spherically symmetric molecules is the Lennard-Jones (LJ) potential, 
\begin{equation}
u\left( r\right) =4\varepsilon \left[ \left( \frac \sigma r\right)
^{12}-\left( \frac \sigma r\right) ^6\right] \,,
\end{equation}
where the parameter $\varepsilon $ is interpreted as the depth of the
potential well and $\sigma $ is the molecular or `collision' diameter. The
reason for choosing this model over other qualitatively similar ones is a
historical one. It is not that it affords a particularly accurate
representation of the potential but rather that it was convenient for
analytical calculations at a time when computers were not available.

The main difficulty in solving the integral equations for $\bar{g}(r)$ and $%
\bar{g}_s(r)$ is immediately apparent. The integrals in Eqs.(\ref{gbar })
and (\ref{gbars}) are not defined if the potential $u(r)$ diverges as $%
r\rightarrow 0$. As a quick remedy we modify the short range behavior and
set $u(r)$ equal to a constant below a certain cutoff distance: $u(r)=u(r_c)$
for $r<r_c$. This arbitrary procedure is justifiable whenever the physical
predictions are not sensitive to the particular value of $r_c$. This is a
serious restriction; we find that insensitivity to the cutoff holds for very
low densities, it fails for even moderately low densities, and under no
circumstances can the treatment be extended to liquid densities. This is not
totally unexpected: behavior at high densities depends crucially on the
short range correlations neglected in MF approximations.

First we discuss the iterative solution of Eqs.(\ref{gbar }) and (\ref{gbars}%
) for $\bar{g}(r)$ and $\bar{g}_s(r)$. We assume the dilute gas
approximation, 
\begin{equation}
g_d(r)=e^{-\beta u(r)},
\end{equation}
as the initial guess that is substituted into the integrals in the right
hand side of Eqs.(\ref{gbar }) and (\ref{gbars}). We find that for
sufficiently low densities the iterative process converges well for both the 
$\bar{g}(r)$ and $\bar{g}_s(r)$ and the two agree. For increasing densities
the solutions develop oscillations of increasingly large amplitude until the
iterations fail to converge.

\begin{figure}[tbp]
\includegraphics*[0in,0in][4in,3in]{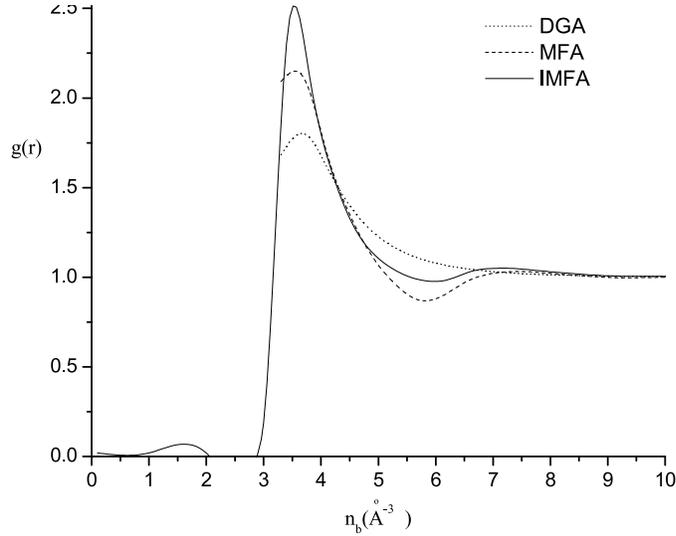}
\caption{ Dilute gas (DGA), mean field (MFA) and improved mean field (IMFA)
approximations for the radial distribution function of Argon.}
\label{comp-g}
\end{figure}

In Fig.\ref{comp-g} we show the dilute gas approximation (DGA) $g_d$, the
mean field approximation (MFA) $\bar{g}$, and the improved mean field
approximation (IMFA) $\bar{g}_s$ for Argon at $n_b=0.00125\,$\AA $^{-3\text{ 
}}$and $T=150.5\,$K. The values used for the LJ parameters are $\varepsilon
=0.0103\,$eV and $\sigma =3.4\,$\AA\ (the choice of $\varepsilon $ and $%
\sigma $ will be discussed below) and the cutoff is set at $r_c=2.9\,$\AA\
(the calculation is insensitive to changes of $r_c$ within a range from
about $2.5$ to $3.1\,$\AA ). The temperature has been chosen sufficiently
low and the density sufficiently high that one is already approaching the
regime where the iterative solutions fail to converge. We find that only
within the restricted region $n_b/kT\lesssim 0.16/\,$\AA $^3$eV the
solutions converge satisfactorily.

The somewhat higher peak of the IMFA can be attributed to the fact that it
partially takes some correlation effects into account. At longer distances
we find that MFA and IMFA are in close agreement.

Next we briefly address the two closely related questions of the
determination of the LJ parameters $\varepsilon $ and $\sigma $ from
experimental data and the overall agreement or not of the IMFA predictions
with experiment. We will use experimental data given in \cite{Expdata} for
the equation of state of Argon and for its internal energy. The equation of
state data are summarized through the coefficients $B\left( T\right) $ and $%
C\left( T\right) $ in the virial expansion

\begin{equation}
\frac P{n_bkT}=1+B\left( T\right) n_b+C\left( T\right) n_b^2+\cdots
\label{virial expansion}
\end{equation}
and data on the internal energy is given in terms of the potential energy
per particle $\langle U\rangle /N$. These data will be compared with the DGA
and IMFA predictions derived from Eqs.(\ref{eq of state}) and (\ref{U per
part}) and with the expression for the second virial coefficient given by

\begin{equation}
B\left( T\right) =\frac{2\pi }{3kT}\int_0^\infty drr^3e^{-\beta u\left(
r\right) }\frac{du\left( r\right) }{dr}.  \label{2ndvirialB}
\end{equation}
The generally accepted values of the LJ parameters are those given in \cite%
{Expdata} by fitting $B(T)$ at $T=273\,$K, 
\begin{equation}
\varepsilon _1=0.01034\,\text{eV}\quad \text{and\quad }\sigma _1=3.405\,%
\text{\AA \thinspace .}  \label{set one}
\end{equation}

\begin{figure}[tbp]
\includegraphics*[0in,0in][4in,3in]{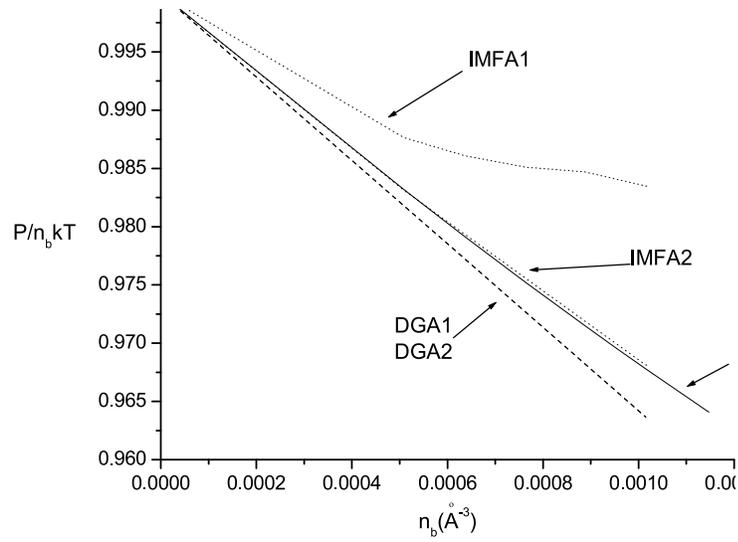}
\caption{ DGA and IMFA equations of state compared with experimental data
for two sets of LJ parameters.}
\label{comp-p}
\end{figure}

In Fig.\ref{comp-p} the DGA and IMFA equations of state for the
parameters $\varepsilon _1$ and $\sigma _1$ are compared with experimental
data. The IMFA curve is in sharp disagreement with the data. Also shown are
the DGA and IMFA predictions for a second set of parameters,

\begin{equation}
\varepsilon _2=0.01032\,\text{eV}\quad \text{and}\quad \sigma _2=3.151\,%
\text{\AA }\,.  \label{set two}
\end{equation}
The DGA curve fits the experimental curve for low density equally well for
either set of parameters, while the IMFA curve for the second set is clearly
the best match with the data. The difference between the $\varepsilon $
parameters is not significant.

\begin{figure}[tbp]
\includegraphics*[0in,0in][4in,3in]{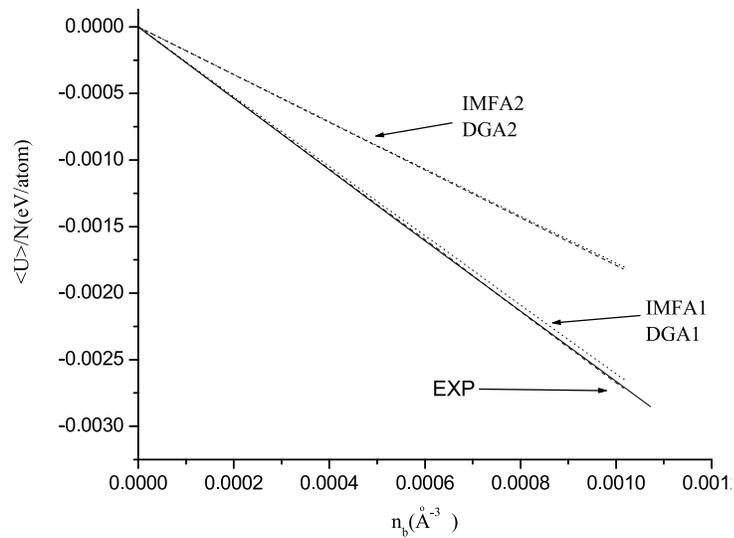}
\caption{ DGA and IMFA potential energy per particle compared with
experimental data for two sets of LJ parameters.}
\label{comp-u}
\end{figure}

On the other hand, as shown in Fig.\ref{comp-u} when the DGA and IMFA
predictions for the potential energy per particle are compared with
experiment it is not the second set of LJ parameters but the first that
yields the better match.

\begin{figure}[tbp]
\includegraphics*[0in,0in][4in,3in]{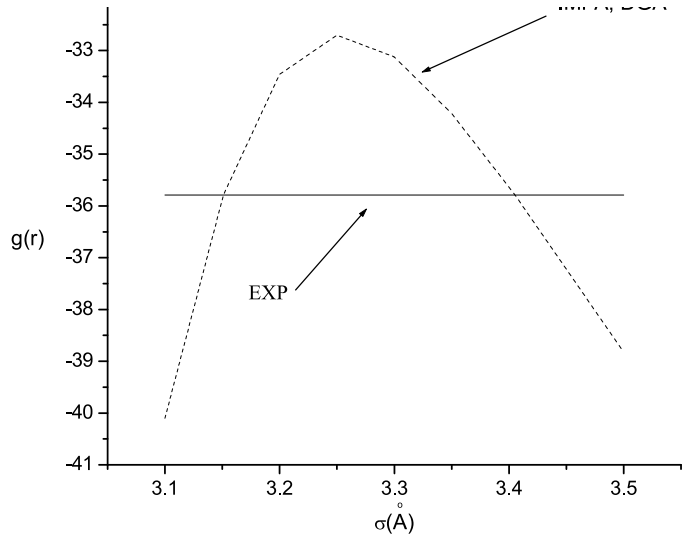}
\caption{ DGA and IMFA predictions for $B$ as a function of $\protect\sigma $
compared to the experimental value.}
\label{Bsol}
\end{figure}

To study this perhaps surprising result from a different perspective we plot
in Fig.\ref{Bsol} the second virial coefficient $B$ at $T=273\,$K for both
the DGA and the IMFA as a function of the collision diameter $\sigma $ for $%
\varepsilon =0.01034\,$eV. Our calculation shows that there are two
solutions for $\sigma $ that match the experimental value, and these two
values are precisely the $\sigma _1$ and $\sigma _2$ in Eqs.(\ref{set one})
and (\ref{set two}).

While it is clear that the IMFA is a definite improvement over the DGA a
number of interesting questions remain unanswered. First, there is the
question of sensitivity of the MF approximations to the short range part of
the potential and the issue of convergence of the iterative solutions. Both
deserve to be studied further. Then, there is the fact that different sets
of LJ parameters are needed to fit different data. Indeed, we found that
different values of $\varepsilon $ and $\sigma $ are needed to fit data at
different temperatures. This can be interpreted as due to intrinsic
inaccuracies in the approximations (DGA, MFA or IMFA) employed or, perhaps
more likely, to the fact that the Lennard-Jones potential is not a good
representation of the true interatomic potential. Finally there are the
inevitable uncertainties in the experimental data which can be considerable %
\cite{Maitland}.

\section{Conclusions}

The method of maximum entropy is traditionally used to assign appropriate
probability distributions. Here we have explored a different use of the
method of maximum entropy, as a technique to generate approximations to
probability distributions that may be intractably difficult to calculate
with. We find that the resulting variational technique includes the
well-known Bogoliuvob variational principle as a special case. The extension
to other generalized canonical ensembles is straightforward and we
explicitly show its application to the density functional formalism and the
derivation of mean field approximations for classical fluids. A side result
of some interest is a simple proof that the density functional formalism is
itself an application of the method of maximum entropy.

Numerical calculations for Argon gas were compared with experimental data.
Just as with other mean field approximations, the particular MF versions
studied here represent improvements over the dilute gas approximation but
remain restricted to very small densities where particle correlations are
small. However, as with so many other variational principles the power of
the method hinges on the right choice of family of trial distributions.
Indeed, following the ME-based variational approach still has the advantage
that the same method can easily be used to generate other approximations
without these shortcomings. If one's interest lies in the physics of denser
fluids the next natural step consists in choosing a family of trial
distributions that provides a better representation of correlation effects.


\begin{thebibliography}{99}
\bibitem{footnote1} The terms `prior' and `posterior' are normally used in
the context of Bayes' theorem; we retain the same terminology because we are
concerned with the similar goal of processing new information to upgrade
from a prior to a posterior. The difference lies in the nature of the
information involved: for Bayes' theorem the information is in the form of
data, for the ME method it is a constraint on the family of allowed
posteriors. The `method of ME' is usually understood in the restricted sense
that one updates from a prior distribution that happens to be uniform. Here
we adopt a broader meaning that includes updates from arbitrary priors and
which involves the maximization of relative entropy.

\bibitem{Jaynes57} E. T. Jaynes, ``Information Theory and Statistical
Mechanics'' Phys. Rev. \textbf{106}, 620 and \textbf{108}, 171 (1957); \emph{%
E. T. Jaynes: Papers on Probability, Statistics and Statistical Physics} ed.
by R. D. Rosenkrantz (Reidel, Dordrecht, 1983).

\bibitem{Shannon48} C. E. Shannon, Bell Systems Tech. Jour. \textbf{27},
379, 623 (1948); C. E. Shannon and W. Weaver, \emph{The Mathematical Theory
of Communication} (Univ. of Illinois Press, Urbana, 1949); N. Wiener, \emph{%
Cybernetics} (MIT Press, Cambridge, 1948); L. Brillouin, \emph{Science and
Information Theory}, (Academic Press, New York, 1956); S. Kullback, \emph{%
Information Theory and Statistics} (Wiley, New York, 1959).

\bibitem{ShoreJohnson80} J. E. Shore and R. W. Johnson, ``Axiomatic
derivation of the Principle of Maximum Entropy and the Principle of Minimum
Cross-Entropy,'' IEEE Trans. Inf. Theory \textbf{IT-26}, 26 (1980); Y.
Tikochinsky, N. Z. Tishby and R. D. Levine, Phys. Rev. Lett. \textbf{52},
1357 (1984) and Phys. Rev. \textbf{A30}, 2638 (1984).

\bibitem{Skilling88} J. Skilling, ``The Axioms of Maximum Entropy'' in \emph{%
Maximum-Entropy and Bayesian Methods in Science and Engineering}, G. J.
Erickson and C. R. Smith (eds.) (Kluwer, Dordrecht, 1988).

\bibitem{Caticha00} A. Caticha, ``Maximum Entropy, Fluctuations and
Priors,'' in \emph{Maximum Entropy and Bayesian Methods in Science and
Engineering}, ed. by A. Mohammad-Djafari, AIP Conf. Proc. \textbf{568}, 94
(2001) (online at arXiv.org/abs/math-ph/0008017).

\bibitem{Callen85} H. B. Callen, \emph{Thermodynamics and an Introduction to
Thermostatistics} (Wiley, New York, 1985).

\bibitem{Opper01} M. Opper and O. Winther, ``From naive mean field theory to
the TAP\ equations'' in \emph{Advanced Mean Field Theory Methods} ed. by M.
Opper and D. Sarad (MIT Press, Cambridge, 2001).

\bibitem{Hansen86} J. P. Hansen and I. R. McDonald, \emph{Theory of Simple
Liquids} (Academic Press, London 1986); C. A. Croxton, \emph{Introduction to
Liquid State Physics} (Wiley, London 1975).

\bibitem{Hohenberg64} P. Hohenberg and W. Kohn, Phys. Rev. \textbf{136},
B864 (1964); N. D. Mermin, Phys. Rev. \textbf{137}, A1441 (1965).

\bibitem{Ebner76} C. Ebner, W. F. Saam and D. Stroud, Phys. Rev. \textbf{A14}%
, 2264 (1976); W. F. Saam and C. Ebner, Phys. Rev. \textbf{A15}, 2566
(1977); G. L. Jones and Soon-Chul Kim, J. Stat. Phys. \textbf{5/6}, 709
(1989).

\bibitem{Evans92} R. Evans, ``Density Functional in the Theory of Nonuniform
Fluids'' in \emph{Fundamentals of Inhomogeneous Fluids} ed. by D. Henderson
(M. Dekker, New York 1992); N. W. Ashcroft, ``Inhomogeneous fluids and the
freezing transition'' in \emph{Density Functional Theory }ed. by E. Gross
and R. Dreizler (NATO ASI vol. \textbf{B337}, Plenum, New York 1995).

\bibitem{Argaman00} N. Argaman and G. Makov, Am. J. Phys. \textbf{68}, 69
(2000); R. Fukuda et al., Prog. Theor. Phys. \textbf{92}, 833 (1994).

\bibitem{Percus62} J. K. Percus, Phys. Rev. Lett. \textbf{8}, 462 (1962).

\bibitem{Expdata} A. Michels, H. Wijker and H. Wijker, Physica \textbf{15},
627 (1949); A. Michels, J. M. Levelt and G. J. Wolkers, Physica \textbf{24},
659 and 769 (1958).

\bibitem{Maitland} G. C. Maitland, M. Rigby and E. B. Smith, \emph{%
Intermolecular Forces -- Their origin and Determination} (Oxford, 1987).
\end{thebibliography}
\end{document}